\documentclass[twocolumn,english,prl]{revtex4}
\usepackage[T1]{fontenc}
\usepackage[latin9]{inputenc}
\usepackage{amsmath}
\usepackage{graphicx}
\usepackage{amssymb}

\makeatletter

\providecommand{\tabularnewline}{\\}

\makeatother

\usepackage{babel}

\begin{document}

\title{Aspects of the confinement mechanism in Landau gauge QCD}

\author{Kai Schwenzer}

\address{Institut für Physik, Karl-Franzens Universität Graz, Universitätsplatz
5, 8010 Graz, Austria}

\begin{abstract}
I analyze the IR fixed point structure of Landau gauge QCD. Precisely
the fixed point with a strong kinematic singularity of the quark-gluon
vertex that proved crucial for the recently proposed confinement mechanism
in the quenched approximation is absent in dynamical QCD.  Therefore,
the IR singularities do not induce asymptotic quark confinement but
the long-range interaction is screened by unquenching loops at scales
 of the order of the quark mass. This provides the prerequisite
for a microscopic description of deconfinement and string breaking.
The fixed points determine the qualitative form of the heavy quark
potential and may be relevant for hot and dense matter.
\end{abstract}
\maketitle
\begin{figure}[t]
\flushleft\includegraphics[scale=0.42]{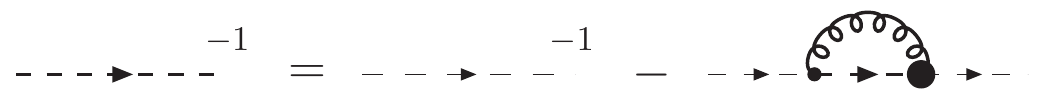}\\
\vspace*{0.2cm}\includegraphics[scale=0.42]{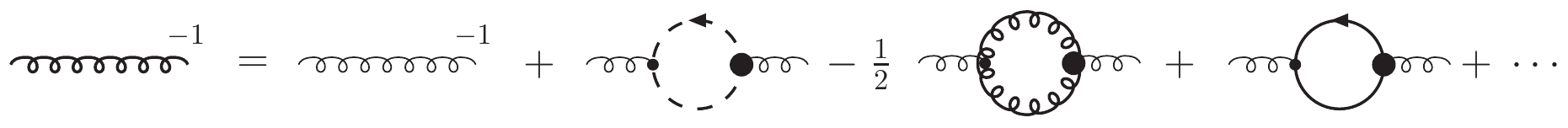}\\
\vspace*{0.1cm}\includegraphics[scale=0.42]{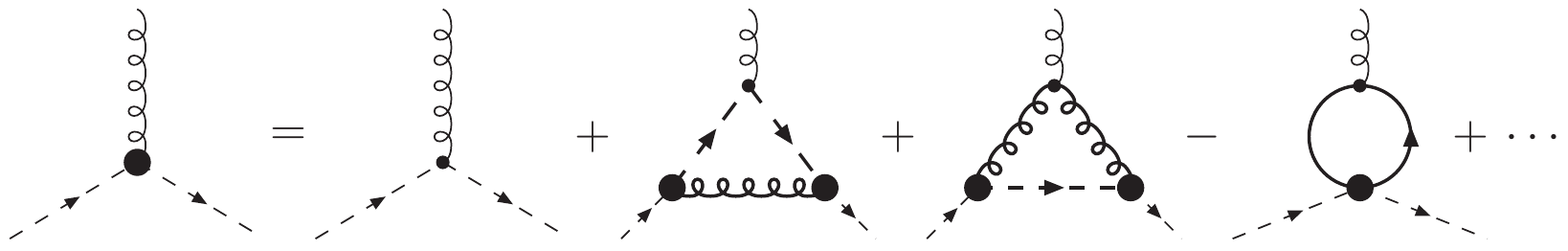}\\
\vspace*{0.1cm}\includegraphics[scale=0.42]{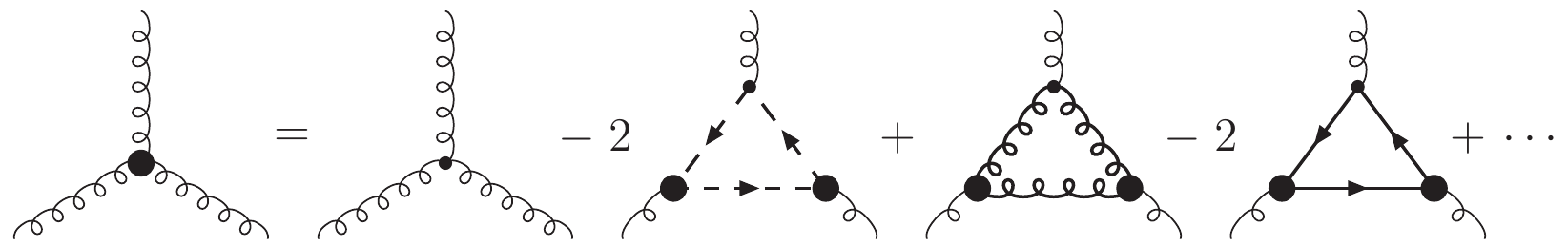}

\caption{The gauge part of the considered coupled DSE system. Ellipses denote
graphs involving neglected 4- and 5-point functions and closed quark
loops are absent in quenched QCD.\label{fig:gauge-part}}

\end{figure}
\begin{figure}[t]
\flushleft\includegraphics[scale=0.42]{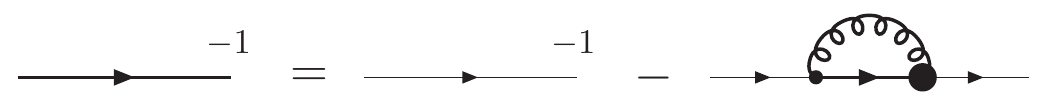}\\
\vspace*{0.1cm}\includegraphics[scale=0.42]{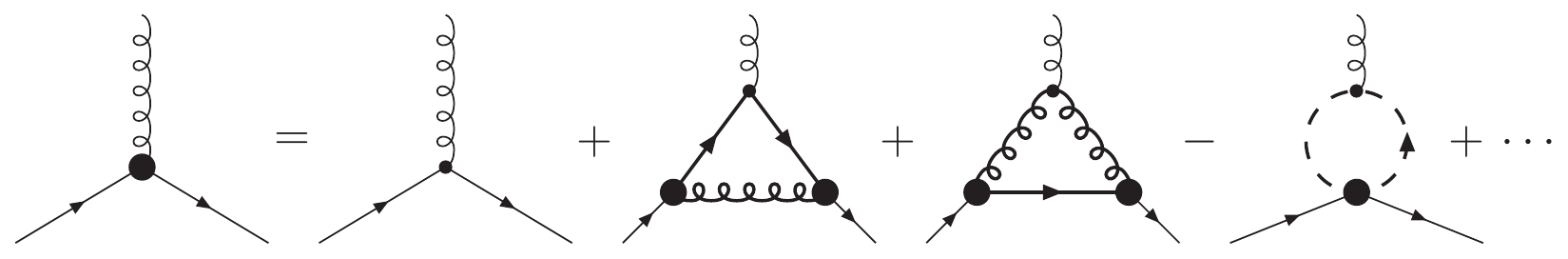}\\
\vspace*{0.1cm}\includegraphics[scale=0.42]{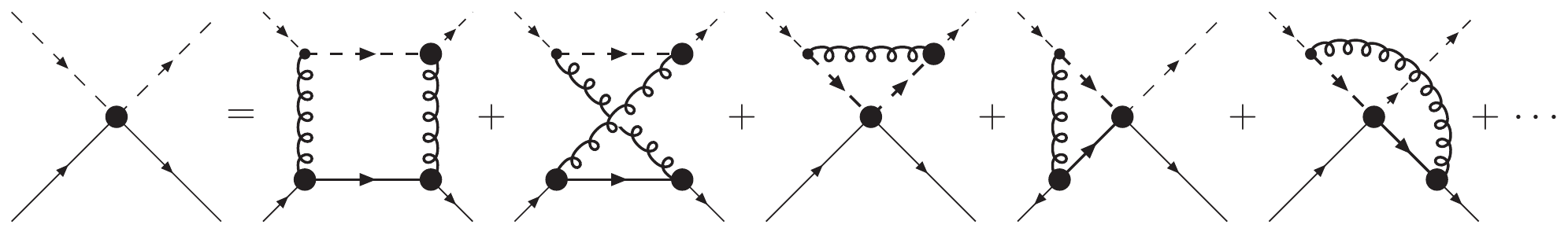}

\caption{The matter part of the considered DSE system.\label{fig:quark-DSEs}}

\end{figure}

Quark confinement is a manifestly non-perturbative phenomenon which
is inherently scale dependent and features several crucial aspects
that a possible mechanism has to explain. In order to give such an
explicit dynamical confinement mechanism in terms of the underlying
colored degrees of freedom this entails to actually specify them and
correspondingly to fix a gauge. In the absence of dynamical quarks
the most direct signature for confinement is an area law of large
Wilson loops \cite{Wilson:1974sk} that can be directly related to
a linear rising potential between static color sources. In momentum
space Green's functions this long-range physics is encoded in the
infrared (IR) regime. The IR divergences of Green's functions are
expected to offer an explanation for confinement since the early work
of Weinberg \cite{Weinberg:1973un}. In Landau gauge Mandelstam \cite{Mandelstam:1979xd}
provided such a mechanism in a simplified approximation to the gluon
Dyson-Schwinger equation (DSE) leading to a strongly divergent gluon
propagator that scales $\sim\!1/p^{4}$ in the IR limit. Yet, it has
been shown recently that an IR enhanced gluon propagator is incompatible
with the DSEs for the gluonic vertices \cite{Alkofer:2008jy}. Instead
there is a consistent IR scaling fixed point of Yang-Mills theory
(YMT) where the ghost sector of the theory dominates in the IR limit
whereas the gluon dynamics is suppressed \cite{vonSmekal:1997is,Lerche:2002ep}.
This scaling solution provides a mechanism for gluon confinement within
the scenarios of Kugo-Ojima \cite{Kugo:1979gm} and Gribov-Zwanziger
\cite{Gribov:1977wm,Zwanziger:1991gz}. It extends to arbitrary
Green's functions \cite{Alkofer:2004it} and kinematics \cite{Alkofer:2008jy},
represents the unique IR scaling fixed point \cite{Fischer:2006vf,Alkofer:2008jy}
and is also obtained with other functional methods \cite{Fischer:2006vf}.
However, it posed a puzzle how suppressed gluons can provide the
strong interaction between color sources seen in lattice QCD \cite{Wilson:1974sk}.\\
Recently it was shown in the quenched approximation that this fixed
point of Landau gauge YMT indeed induces a strong IR quark interaction
that leads to a linear, confining potential between quarks \cite{Alkofer:2008tt}.
This confinement mechanism relies on a self-consistently enhanced
kinematic divergence of the quark-gluon vertex in the soft-gluon limit
that the external gluon momentum vanishes. The question is then
how unquenching effects can circumvent this static confinement and
thereby account for string breaking and hadronization seen in collider
experiments. Such a mechanism must be related to scales of the order
of the masses of created hadrons. In this work I outline a comprehensive
analysis of the IR fixed point structure of Landau gauge QCD in the
unquenched case. It confirms the screening of gluonic interactions
via quark loops in the on-shell region that could  {}``break the
gluonic string'' and drive the hadronization process.\\
The non-perturbative dynamics of QCD is described by the system of
DSEs which is generated algorithmically \cite{Alkofer:2008nt}.
The DSEs form an infinitely coupled tower of equations that in general
requires approximations. Yet, it can be shown \cite{long-paper}
that the equations for all $n$-point functions with $n\!>\!4$ are
linear in the sense that the corresponding $n$-point function appears
at most once in each loop graph. As far as the IR singularities are
concerned there is thereby no self-consistent enhancement mechanism
in these equations. They also include no tree level term and are hence
effectively determined by lower order Green's functions. Therefore,
any IR enhancement has to be generated by the primitively divergent
Green's functions and the truncation can be restricted to the latter.
Moreover, it has been found previously \cite{Alkofer:2008jy,Alkofer:2004it}
that all contributions including a 4-gluon vertex are strongly IR
suppressed so that this vertex can be neglected as well. However,
the DSE system features a peculiarity compared to other functional
approaches like the functional renormalization group (FRG) \cite{Fischer:2006vf}
or $n$-particle irreducible ($n$PI) actions \cite{Berges:2004pu}
in that it involves a bare vertex in every DSE so that IR strength
can be missing in the lowest order diagrams when they involve enhanced
vertices. Since ghosts and quarks do not couple at the tree level
this requires to include the ghost-quark vertex to consider all
IR leading terms. The considered truncation is then given by the DSEs
in the gauge and matter sector in figs. \ref{fig:gauge-part} and
\ref{fig:quark-DSEs}.\\
For the analysis of the IR regime a flexible power counting scheme
\cite{Lerche:2002ep,Alkofer:2004it,Alkofer:2008jy} has been developed
that yields the IR fixed point structure without an actual solution
of the non-linear integral equations.  Here I only sketch the main
steps of the analysis which is similar to the one previously performed
in the gauge sector \cite{Alkofer:2008jy} and extends the one in
the quenched approximation \cite{Alkofer:2008tt}. The details will
be presented in a forthcoming article \cite{long-paper}. The infrared
analysis relies on the observation that YMT is scale invariant at
the classical level. Quantum fluctuations induce a dynamical scale
$\Lambda_{\mathrm{QCD}}$ and in QCD the matter sector introduces
in addition explicit mass scales, but it is experimentally established
that strong interaction has no scales far below the MeV scale. Correspondingly,
it is a reasonable assumption that in the IR regime $p\!\ll\!\Lambda_{\mathrm{QCD}}$
the scale dependence of all Green's functions is given by a power
law scaling in terms of the the dimensionless ratio $\rho\!=\! p/\Lambda_{\mathrm{QCD}}$,
parametrized for the propagators by\[
D_{\mu\nu}^{ab}\left(p\right)\sim\left(\rho^{2}\right)^{\delta_{gl}-1}\!,\; G^{ab}\left(p\right)\sim\left(\rho^{2}\right)^{\delta_{gh}-1}\!,\; S\left(p\right)\sim\left(\rho^{2}\right)^{\delta_{q}}\]
where $\delta_{gl}$, $\delta_{gh}$ and $\delta_{q}$ denote the
anomalous IR exponents, distinguished by lower indices. The vertices
depend on several external momenta and can feature divergences in
various kinematic configurations that are described by different IR
exponents \cite{Alkofer:2008jy,Alkofer:2008tt}, distinguished by
upper indices. In the uniform limit that all external momenta vanish
these exponents are $\delta_{gg}^{u}$ for the ghost-gluon vertex,
$\delta_{3g}^{u}$ for the 3-gluon vertex, $\delta_{qg}^{u}$ for
the quark-gluon vertex and $\delta_{qgh}^{u}$ for the quark-ghost
vertex. In addition there are exponents $\delta_{qgh}^{2gh}$, $\delta_{qgh}^{2q}$,
$\delta_{qgh}^{qgh}$ when two external momenta vanish, $\delta_{gg}^{gh}$,
$\delta_{gg}^{gl},$ $\delta_{3g}^{gl}$, $\delta_{qg}^{q}$, $\delta_{qg}^{gl}$,
$\delta_{qgh}^{gh}$, $\delta_{qgh}^{q}$ when a single external momentum
vanishes and finally $\delta_{qgh}^{st}$ when only the difference
of the two external quark respectively ghost legs is soft. Where there
are different tensor structures these are the exponents of the IR
leading part(s). \\
The loop integrals in the DSEs are dominated by the singularities
of the integrand and receive contributions from loop momenta in the
vicinity of all external momentum and mass scales \cite{Alkofer:2008jy},
i.e. in QCD also by scales of the order of the constituent quark masses.
In the IR limit there is a clear scale separation and one can divide
the contributions from soft and hard loop momenta. By determining
the IR scaling of these distinct contributions from the different
loop graphs via a power counting analysis the DSE system translates
to a coupled set of algebraic equations for the anomalous exponents.
In each DSE the IR leading term on the right hand side, corresponding
to the smallest IR exponent, determines the IR behavior of the Green's
function. When hard momenta are present there are various possible
cancellations in the IR analysis that affect the canonical scaling
of Green's functions. These arise from gluon transversality, the symmetry
of Green's functions \cite{Alkofer:2008jy} and from the Dirac structure
\cite{Alkofer:2008tt}. There is an additional suppression of the
contribution from hard loop momenta $k$ where the quark propagator
can be expanded in the soft external momentum $p$\[
\frac{Z\!\!\left(\!\left(k\!+\! p\right)^{2}\!\right)}{i\!\left(\slash\!\!\! k\!+\!\slash\!\!\! p\right)\!+\! M\!\!\left(\!\left(k\!+\! p\right)^{2}\!\right)}\!=\!\frac{Z\!\left(k^{2}\right)}{i\slash\!\!\! k\!+\! M\!\left(k^{2}\right)}\!\!\left(\!1\!-\!\frac{2\, p\!\cdot\! k}{k^{2}\!+\! M^{2}\!\left(k^{2}\right)}\!+\!\cdots\!\right)\]
which shows that if there is no constituent quark mass $m_{q}\!\equiv\! M\!\left(0\right)\!=\!0$
or another hard scale in the integral the leading term is scale-free
and is removed in the renormalization procedure so that the propagator
is actually suppressed in $p$. The gauge propagators feature an analogous
suppression by $p^{2}$ \cite{Alkofer:2008jy}. In the power counting
analysis this is considered by symbols that take values $\mu_{q}\!=\!\tfrac{1}{2}$
and $\mu_{gl}\!=\!1$ when the propagators are massless, i.e. $\delta_{q}\!=\!-\,\tfrac{1}{2}$
and $\delta_{gl}\!=\!0$, and which vanish otherwise. Analogous symbols
$m_{q}^{0}$ and $\mu_{q}^{0}$ are introduced at the tree level.\\
The power counting analysis for the DSE system figs. \ref{fig:gauge-part}
and \ref{fig:quark-DSEs} results in an extensive system of coupled
algebraic equations for the above 18 IR exponents \cite{long-paper}.
Fortunately, there are important constraints on the IR exponents,
given in \cite{Alkofer:2008jy}, and analogous ones in the quark
sector that simplify the system considerably. Previously such constraints
had been obtained from the assumption that a skeleton expansion of
the vertices should not explicitly diverge. Yet, the same constraints
are obtained independently from the complementary set of FRG equations
without additional assumptions \cite{Markus,Fischer:2006vf}. Since
the equations for the quark-ghost vertex are linear they can be solved
by successive replacements of the possible solutions and the connection
between the exponents in different kinematic limits provides further
complementary constraints \cite{long-paper}. Applying the constraints
yields that the fermionic 3-point vertices are not divergent when
only a fermion momentum vanishes $\delta_{gg}^{gh}\!=\!\delta_{qg}^{q}\!=\!0$
and leaves the simplified system for the IR exponents of the propagators\begin{align}
1\!-\!\delta_{gh}= & \min\!\left(1;\delta_{gg}^{u}\!+\!\delta_{gh}\!+\!\delta_{gl}\!+\!1\right)\label{eq:ghost-prop}\\
-\delta_{q}= & \min\!\left(\mu_{q}^{0};\delta_{qg}^{u}\!+\!\delta_{q}\!+\!\delta_{gl}\!+\!1,\mu_{q}\right)\label{eq:quark-prop}\\
1\!-\!\delta_{gl}= & \min\!\left(1;\delta_{gg}^{u}\!+\!2\delta_{gh}\!+\!1,\delta_{gg}^{gl}\!+\!1;\delta_{3g}^{gl}\!+\!\mu_{gl};\right.\label{eq:gluon-prop}\\
 & \!\left.\left[\delta_{qg}^{u}\!+\!2\delta_{q}\!+\!2,\delta_{qg}^{gl}\!+\!\mu_{q}\right]\right)\nonumber \end{align}
and of the remaining primitively divergent vertices\begin{align}
\delta_{gg}^{u}\!+\!\tfrac{1}{2}= & \min\!\left(\tfrac{1}{2};\cdots;2\delta_{gg}^{u}\!+\!\delta_{gh}\!+\!2\delta_{gl}\!+\!\tfrac{1}{2};\right.\label{eq:ghost-vertex-uni}\\
 & \!\left.\left[2\delta_{q}\!+\!3\!+\!\mu_{gl}\,\mu_{q},\delta_{gg}^{u}\!+\!2\delta_{qg}^{gl}\!+\!\delta_{gh}\!+\!2\delta_{gl}\!+\!\tfrac{1}{2}\!+\!\mu_{q}\right]\right)\nonumber \\
\delta_{gg}^{gl}= & \min\!\left(0;2\delta_{gh}\!+\!\tfrac{3}{2};2\delta_{gg}^{gl}\!+\!2\delta_{gl}\!+\!\tfrac{1}{2};\right.\label{eq:gluon-vertex-soft}\\
 & \!\left.\left[\delta_{gg}^{gl}\!+\!2\delta_{qg}^{gl}\!+\!2\delta_{gl},\delta_{gg}^{u}\!+\!2\delta_{qg}^{gl}\!+\!3\delta_{gh}\!+\!2\delta_{gl}\!+\!1\right]\right)\nonumber \\
\delta_{3g}^{u}\!+\!\tfrac{1}{2}= & \min\!\left(\tfrac{1}{2};2\delta_{gg}^{u}\!+\!3\delta_{gh}\!+\!\tfrac{1}{2},2\delta_{gg}^{gl}\!+\!1;2\delta_{3g}^{gl}\!+\!\tfrac{1}{2}\!+\!\mu_{gl};\right.\nonumber \\
 & \!\left.\left[2\delta_{qg}^{u}\!+\!3\delta_{q}\!+\!2,2\delta_{qg}^{gl}\!+\!\tfrac{1}{2}\!+\!\mu_{q}\right]\right)\label{eq:gluon-vertex-uni}\\
\delta_{3g}^{gl}= & \min\!\left(0;2\delta_{gh}\!+\!1;2\delta_{3g}^{gl}\!+\!2\delta_{gl}\!+\!\tfrac{1}{2};\left[2\delta_{q}\!+\!2\right]\right)\label{eq:gluon-vertex-soft}\\
\delta_{qg}^{u}= & \min\!\left(0;\cdots;\delta_{gg}^{u}\!+\!2\delta_{qg}^{u}\!+\!3\delta_{gh}\!+\!\delta_{q}\!+\!2\delta_{gl}\!+\!\tfrac{1}{2},\right.\label{eq:quark-vertex-uni}\\
 & \!\left.2\delta_{gh}\!+\!\tfrac{3}{2}\!+\!\mu_{gl}\,\mu_{q},\delta_{gg}^{gl}\!+\!2\delta_{qg}^{u}\!+\!\delta_{q}\!+\!2\delta_{gl}\!+\!1\right)\nonumber \\
\delta_{qg}^{gl}= & \min\!\left(0;\cdots;\delta_{gg}^{u}\!+\!2\delta_{qg}^{gl}\!+\!3\delta_{gh}\!+\!2\delta_{gl}\!+\!\tfrac{1}{2},\right.\label{eq:quark-vertex-soft}\\
 & \!\left.2\delta_{gh}\!+\!\tfrac{3}{2},\delta_{gg}^{gl}\!+\!2\delta_{qg}^{gl}\!+\!2\delta_{gl}\right)\nonumber \end{align}
where the contributions from different graphs in figs. \ref{fig:gauge-part}
and \ref{fig:quark-DSEs} are separated by semicolons and contributions
from different regions of the same loop integral by commas. It
is instructive to recall the solution of the quenched system \cite{Alkofer:2008tt}
first, where the terms in brackets arising from closed quark loops
are absent. There the quark-gluon vertex features different solutions
with a trivial respectively a strongly enhanced IR scaling. In the
DSEs the latter arises from the IR region of the ghost loop diagram
in fig. \ref{fig:quark-DSEs} corresponding to the first non-trivial
term in eqs. (\ref{eq:quark-vertex-uni}) and (\ref{eq:quark-vertex-soft}).
Inserting the dressed quark-ghost vertex, results in a non-Abelian
graph with a ghost-triangle, representing the IR leading contribution
\cite{Alkofer:2008jy} to the 3-gluon vertex in eq. (\ref{eq:gluon-vertex-uni}).
Since such a fully dressed non-Abelian graph arises directly both
in the FRG \cite{Fischer:2006vf} and $n$PI actions \cite{Berges:2004pu},
the dynamics seems to be represented more efficiently in these approaches.
More importantly, for the strong soft-gluon singularity the unquenching
contributions in the last term in brackets in eqs. (\ref{eq:gluon-prop})
and (\ref{eq:gluon-vertex-uni}) are incompatible with the gauge
sector, e.g. $1\!-\!\delta_{gl}\nleq\delta_{qg}^{gl}$.\\
The starting point for the full solution of the unquenched system
is eq. (\ref{eq:ghost-prop}) for the ghost propagator. As it stands
the bare term is leading and in addition to the trivial, perturbative
solution that is also realized in the ultraviolet (UV) regime one
obtains a decoupling solution with a bare ghost and an IR finite gluon
propagator \cite{Boucaud:2008ji,Fischer:2008uz}. Due to the suppressed
gluonic interaction there are no IR enhanced Green's functions.
The decoupling solution seems to be found in current lattice studies
\cite{Bowman:2007du}. Yet the study of the IR regime of a gauge
theory on the lattice is a notoriously hard problem and there are
several issues like the present implementation of the gauge condition
and the related contamination with Gribov copies that demand caution
\cite{Fischer:2008uz}. In particular, since there is another solution
that explains important qualitative aspects of strong interaction
physics better: For an IR enhanced propagator the renormalization
procedure can also be performed in a way that the tree level term
in the ghost equation is cancelled identically. This yields the scaling
solution where $\delta_{gl}\!=\!-2\delta_{gh}\,\equiv\!2\kappa$ \cite{vonSmekal:1997is},
depending on a parameter $\kappa\!\geq\!0$ since the gluon equation
(\ref{eq:gluon-prop}) is then identically fulfilled by the ghost-loop
exponent \cite{Alkofer:2008jy}. The correspondingly subleading contribution
from unquenching loops with hard momenta in eq. (\ref{eq:gluon-prop})
yields the important constraint $\delta_{qg}^{gl}\!\geq\!1\!-\!2\kappa\!-\!\mu_{q}$.
The scaling solution depends on the three qualitatively different
cases of an infinite, a finite or a vanishing constituent quark mass
in eq. (\ref{eq:quark-prop}). The first case is similar to the quenched
limit, but here quarks are entirely static and merely act as sources.
Using the available constraints the solution of the residual system
is possible without further assumptions and the resulting IR fixed
points of QCD are given in table \ref{tab:fixed-points}. Strikingly
for all QCD fixed points the gauge sector is not altered by the quark
dynamics compared to the YMT result \cite{Alkofer:2008jy,Alkofer:2004it}.
The main result of this article is, however, that due to the above
constraint the solution with a strong soft-gluon singularity of the
quark-gluon vertex is ruled out in dynamical QCD and only the solution
with a trivial soft-scaling remains. This happens since hard virtual
quark loops with momenta around the quark mass screen the long-range
gluonic interaction.\\
\begin{table*}

\begin{tabular}{|c|c|c|c|c|c|c|c|c|c|c|c|c|}
\hline 
 &  & $\delta_{gh}$ & $\delta_{gl}$ & $\delta_{q}$ & $\delta_{gg}^{u}$ & $\delta_{gg}^{gl}$ & $\delta_{gg}^{gh}$ & $\delta_{3g}^{u}$ & $\delta_{3g}^{gl}$ & $\delta_{qg}^{u}$ & $\delta_{qg}^{gl}$ & $\delta_{qg}^{q}$\tabularnewline
\hline 
scaling & static ($m_{q}\to\infty$)$\,/\,$quenched & $-\kappa$ & $2\kappa$ & -$\,/\,0$ & $0$ & $0$ & $0$ & $-3\kappa$ & $\min\!\left(0,1\!-\!2\kappa\right)$ & $\,$-$\,/-\!\tfrac{1}{2}\!-\!\kappa\:\vee\:0$ & ${\bf -}{\bf \tfrac{1}{2}}{\bf -}{\bf \kappa}\:\vee\:0$ & $0$\tabularnewline
\hline
 & massive ($m_{q}>0$,$m_{q}^{0}\geq0$) & $-\kappa$ & $2\kappa$ & $0$ & $0$ & $0$ & $0$ & $-3\kappa$ & $\min\!\left(0,1\!-\!2\kappa\right)$ & $-\tfrac{1}{2}\!-\!\kappa\:\vee\:0$ & ${\bf 0}$ & $0$\tabularnewline
\hline
 & chiral ($m_{q}=m_{q}^{0}=0$) & $-\kappa$ & $2\kappa$ & $-\tfrac{1}{2}$ & $0$ & $0$ & $0$ & $-3\kappa$ & $\min\!\left(0,1\!-\!2\kappa\right)$ & ${\bf -}{\bf \kappa}\:\vee\:0$ & ${\bf 0}$ & $0$\tabularnewline
\hline
decoupling  & ($\forall\, m_{q}$) & $0$ & $1$ & $-\tfrac{1}{2}\:\vee\:0$ & $0$ & $0$ & $0$ & $0$ & $0$ & $0$ & ${\bf 0}$ & $0$\tabularnewline
\hline
perturbative & ($\forall\, m_{q}$, both IR \& UV) & $0$ & $0$ & $-\tfrac{1}{2}\:\vee\:0$ & $0$ & $0$ & $0$ & $0$ & $0$ & $0$ & ${\bf 0}$ & $0$\tabularnewline
\hline
\end{tabular}

\caption{The anomalous power law exponents of the leading Green's functions
for the different fixed points of QCD distinguished by the constituent
quark mass $m_{q}$. These power laws are only valid up to logarithmic
corrections. The full vertices include also the canonical scaling
dimension $-1$ for the gauge propagators and $\frac{1}{2}$ for the
gauge 3-point vertices in the uniform limit. The value of $\kappa$
is fixed by an explicit IR solution and the best known value is $\kappa\!\approx\!0.59$
\cite{Lerche:2002ep}. There are additional fixed points in the chiral
case where the quark propagator is IR divergent \cite{long-paper},
yet physically they are excluded by the current quark masses.\label{tab:fixed-points}}
\end{table*}%
\begin{figure}
\begin{minipage}[c][1\totalheight]{0.6\columnwidth}%
\flushleft\includegraphics[scale=0.42]{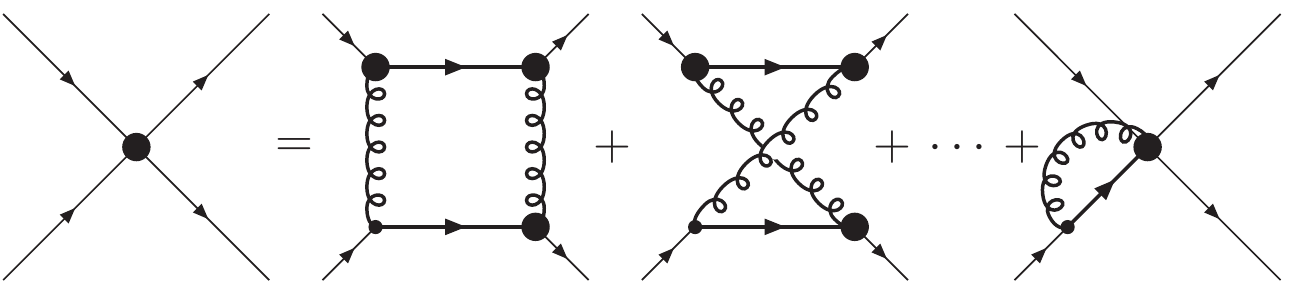}\\
\vspace*{0.1cm}\includegraphics[scale=0.42]{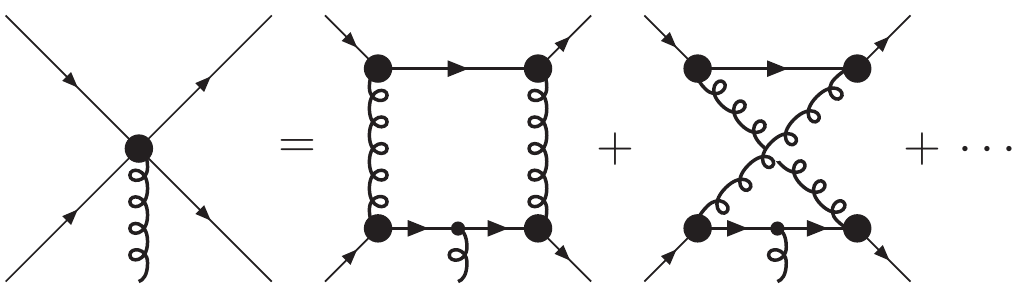}%
\end{minipage}%
\begin{minipage}[c][1\totalheight]{0.4\columnwidth}%
\flushright\includegraphics[scale=0.32]{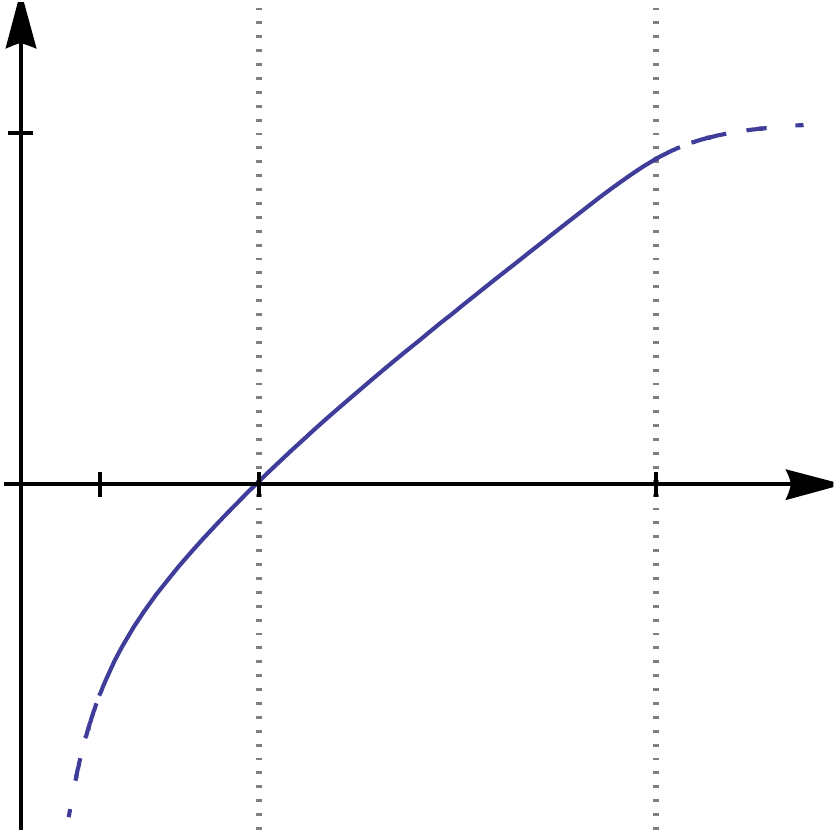}

\flushleft \vspace*{-3.15cm} \hspace*{0.8cm} $V$ \flushleft \vspace*{-0.45cm} \hspace*{0.15cm} $\scriptstyle 2m_q$
\flushleft \vspace*{-0.3cm} \hspace*{0.975cm} ${\mathcal P}$ \hspace*{0.575cm} $\mathcal S$ \hspace*{0.55cm} $\mathcal M$
\flushleft \vspace*{-0.05cm} \hspace*{3.2cm} $x$
\flushleft \vspace*{-0.2cm} \hspace*{0.7cm} $\scriptstyle \frac{1}{m_q}$ \hspace*{0.15cm} $\scriptstyle \frac{1} {\Lambda_{\mathrm{QCD}}}$ \hspace*{0.3cm} $\scriptstyle \frac{2m_q}{\sigma}$ \vspace*{0.3cm}%
\end{minipage}

\caption{\emph{left}: The DSE for the 4-quark vertex and IR leading contributions
to the arising 4-quark-gluon vertex\label{fig:4-quark-DSE}, \emph{right}:
 Heavy quark potential with areas of application of the perturbative
($\mathcal{P}$) UV, static ($\mathcal{S}$) and massive ($\mathcal{M}$)
IR fixed points.}

\end{figure}
Now let us discuss the impact of the IR divergences on quark confinement
which should in the case of mesons be described by the 4-quark vertex
DSE given in the first line of fig. \ref{fig:4-quark-DSE}. The quarks
in hadrons have momenta of the order of hundreds of MeV and correspondingly
the relevant kinematics for confinement is given by hard external
quark momenta. Though the exchanged gluon momentum becomes soft when
they are far spatially separated. Using the IR exponents in the static
limit given in table \ref{tab:fixed-points}, the first diagrams
in fig. \ref{fig:4-quark-DSE} do not induce strong long-range interactions
since IR strength is missing due to the bare quark-gluon vertex \cite{Alkofer:2008tt}.
However, this is not the case for the last graph with the higher order
4-quark-gluon vertex. Its DSE in the second line of fig. \ref{fig:4-quark-DSE}
includes graphs where all four quark-gluon vertices attached to the
exchanged soft gluons are dressed. The kinematic regime where the
small loop in the first line is dominated by momenta of the order
of the quark mass and the one in the second line is soft yields a
far stronger divergence. In the heavy mass limit the hard loop shrinks
to a point and the leading corrections reduce to graphs similar to
the first two in the 4-quark DSE but with all four quark-gluon vertices
dressed - again in complete agreement with the FRG \cite{Fischer:2006vf}
or nPI methods \cite{Berges:2004pu}. With the strong soft-gluon
singularity of the dressed quark-gluon vertices this yields an IR
interaction that scales $\sim\!1/p^{4}$ corresponding to a linear,
confining potential in coordinate space \cite{Alkofer:2008tt}.\\
In contrast, in the unquenched case the quark-gluon vertex is not
enhanced corresponding to a 4-quark interaction $\sim\! p^{8\kappa}$
which yields a strongly suppressed long-range behavior and cannot
be interpreted by a non-relativistic potential. Instead, the long-range
interaction, found in the static respectively quenched case, is screened
by virtual quark loops at scales of the order of the quark mass and
dynamical QCD is indeed not asymptotically confining. Yet, in the
heavy mass limit $m_{q}\!\gg\!\Lambda_{\mathrm{QCD}}$ and for momenta
$p\!\ll\! m_{q}$ where the quenched limit is a valid approximation
one can expect a linear potential over distances for which $V\!\sim\!\sigma x\!\ll\!2m_{q}$.
In the short distance regime $p\!\sim\!1/x\!\gg\!\Lambda_{\mathrm{QCD}}$
the perturbative UV fixed point is realized which yields an interaction
$\sim\!1/p^{2}$ corresponding to a Coulomb potential, whereas the
potential picture breaks down for distances $x\!\lesssim\!1/m_{q}$.
Correspondingly the qualitative form of the heavy quark potential
is entirely determined by the fixed points of QCD as shown schematically
in fig. \ref{fig:4-quark-DSE}. The arising screening scale is precisely
of the order of the masses of corresponding mesonic states and thereby
it is not hard to imagine that {}``the string breaks'' and real
particles are produced when the system is sufficiently excited.
This is described by higher Green's functions, like the 6-quark vertex,
and it remains to be shown that only color singlets are produced.
However, due to the intermediate rise in fig. \ref{fig:4-quark-DSE}
it is clear that produced quarks are confined into bound states when
their excitation energy is small enough. The presence of light quarks
surely goes beyond the present IR analysis and requires computations
at finite hadronic scales.\\
Finally the above fixed points could be relevant for hot and dense
matter. In the limit of asymptotic temperature $T$ and/or chemical
potential $\mu$ the quark masses become irrelevant in comparison
and the chiral cases in table \ref{tab:fixed-points}, which do not
induce quark confinement, may be relevant for some scale region.
Yet, in addition to the perturbative fixed point, describing complete
deconfinement, the chiral scaling fixed points feature a confined
gauge sector. The stronger one exhibits a Coulombic quark interaction
as in the perturbative case and may be relevant for the strongly coupled
plasma observed in heavy ion collisions. In contrast, the weaker one
corresponds to a  quark sector that completely decouples from a strongly
coupled gauge sector in the IR limit. Interestingly, such a behavior
has been independently found in an analysis of the low energy dynamics
of ungapped modes  at high $\mu$ \cite{Schafer:2005mc}.

\begin{acknowledgments}
I am grateful to R. Alkofer, C. Fischer, M. Huber, F. Llanes-Estrada,
A. Maas \& J. Pawlowski for helpful discussions and the FWF for support
via grant M979-N16.
\end{acknowledgments}


\begin{thebibliography}{10}
\bibitem{Wilson:1974sk}K.~G.~Wilson, Phys.\ Rev.\  D {\bf 10} (1974) 2445.

\bibitem{Weinberg:1973un}S.~Weinberg, Phys.\ Rev.\ Lett.\  {\bf 31} (1973) 494.

\bibitem{Mandelstam:1979xd}S.~Mandelstam, Phys.\ Rev.\ D {\bf 20} (1979) 3223.

\bibitem{Alkofer:2008jy}R.~Alkofer, M.~Q.~Huber, K.~Schwenzer, arXiv:0801.2762. 

\bibitem{vonSmekal:1997is}L.~von Smekal, R.~Alkofer and A.~Hauck, Phys.\ Rev.\ Lett.\  {\bf 79} (1997) 3591;
Annals Phys.\  {\bf 267} (1998) 1.

\bibitem{Lerche:2002ep}C.~Lerche, L.~von Smekal, Phys.\ Rev.\ D {\bf 65} (2002) 125006.

\bibitem{Kugo:1979gm}T.~Kugo, I.~Ojima,\! Prog.\! Theor.\! Phys.\! Suppl.\ {\bf 66} (1979) 1.

\bibitem{Gribov:1977wm}V.~N.~Gribov, Nucl.\ Phys.\ B {\bf 139} (1978) 1.

\bibitem{Zwanziger:1991gz}D.~Zwanziger, Nucl.\ Phys.\  B {\bf 364} (1991) 127.

\bibitem{Alkofer:2004it}R.~Alkofer, C.~S.~Fischer, F.~J.~Llanes-Estrada, \!Phys.\! Lett.\! B {\bf 611} (2005) 279; 
\! Mod.\! Phys.\! Lett.\!  A {\bf 23} (2008) 1105.

\bibitem{Fischer:2006vf}C.~S.~Fischer and J.~M.~Pawlowski,\ Phys.\ Rev.\ D {\bf 75} (2007) 025012; J.~M.~Pawlowski, D.~F.~Litim, S.~Nedelko and L.~von Smekal, Phys.\ Rev.\ Lett.\  {\bf 93}, 152002 (2004).

\bibitem{Alkofer:2008tt}R.~Alkofer, C.~S.~Fischer, F.~J.~Llanes-Estrada and K.~Schwenzer, arXiv:0804.3042; 
Annals Phys. in press.

\bibitem{Alkofer:2008nt}R.~Alkofer, M.~Q.~Huber, K.~Schwenzer, arXiv:0808.2939. 

\bibitem{long-paper}K.~Schwenzer {\it et al.}, in preparation.

\bibitem{Berges:2004pu}J.~Berges, Phys.\ Rev.\ D {\bf 70} (2004) 105010.

\bibitem{Markus}M.~Q.~Huber, 
R.~Alkofer and K.~Schwenzer, in prep.

\bibitem{Boucaud:2008ji}Ph.~Boucaud, {\it et al.}, JHEP {\bf 0806} (2008) 012.

\bibitem{Fischer:2008uz}C.S.~Fischer, A.~Maas, J.M.~Pawlowski, arXiv:0810.1987. 

\bibitem{Bowman:2007du}P.~O.~Bowman {\it et al.}, Phys.\ Rev.\  D {\bf 76} (2007) 094505;  I.~L.~Bogolubsky
{\it et al.}, Phys.\ Rev.\  D {\bf 77} (2008) 014504; 
A.~Cucchieri, and T.~Mendes, PoS {\bf LAT2007} (2007) 297.

\bibitem{Schafer:2005mc}T.~Schafer and K.~Schwenzer, Phys.\ Rev.\ Lett.\  {\bf 97} (2006) 092301;
K.~Schwenzer, Nucl.\ Phys.\  A {\bf 785} (2007) 241.
\end{thebibliography}
\end{document}